\documentclass[aps,prl,superscriptaddress,showpacs]{revtex4-1}
\usepackage{graphics}
\newcommand{\unit}[2]{\ensuremath {#1}\,{\mathrm{#2}}}
\newcommand{\Fig}[1]{Fig.~\ref{fig:#1}}

\newcommand{\Ip}{\ensuremath I_\mathrm{p}}
\newcommand{\Hqb}{\ensuremath H_{\mathrm{QB}}}
\newcommand{\Htotal}{\ensuremath H_{\mathrm{total}}}

\newcommand{\Tfluct}{\ensuremath T_{\mathrm{TLS}}}
\newcommand{\Tbath}{\ensuremath T_{\mathrm{bath}}}
\newcommand{\Tcav}{\ensuremath T_{\mathrm{res}}}

\newcommand{\sigmazTLS}{\ensuremath \sigma_{\mathrm{z,TLS}}}
\newcommand{\sigmaxTLS}{\ensuremath \sigma_{\mathrm{x,TLS}}}

\newcommand{\sigmamQB}{\ensuremath \sigma_{\mathrm{+,QB}}}
\newcommand{\sigmapQB}{\ensuremath \sigma_{\mathrm{-,QB}}}
\newcommand{\sigmazQB}{\ensuremath \sigma_{\mathrm{z,QB}}}
\newcommand{\sigmaxQB}{\ensuremath \sigma_{\mathrm{x,QB}}}

\newcommand{\ttwoqb}{\ensuremath t_{\mathrm{2,QB}}}

\newcommand{\nuqb}[1][]{\ensuremath \nu_{\mathrm{QB}#1}}

\newcommand{\mod}{\ensuremath {\mathrm{mod}}}

\newcommand{\state}[1]{\ensuremath \left|#1\right>}
\newcommand{\planck}{h}

\newcommand{\nures}{\ensuremath \nu_{\mathrm{res}}}

\newcommand{\lres}{\ensuremath l_{\mathrm{res}}}

\newcommand{\Phiext}{\ensuremath \Phi_{\mathrm{ext}}}
\newcommand{\DeltaPhiext}{\ensuremath \delta\Phi_{\mathrm{ext}}}

\newcommand{\phim}{\ensuremath \phi_{\mathrm{m}}}
\newcommand{\Phizero}{\ensuremath \Phi_{\mathrm{0}}}
\newcommand{\Phicontrol}{\ensuremath \Phi}

\newcommand{\EJ}{\ensuremath E_{\mathrm{J}}}

\newcommand{\nufluct}{\ensuremath \nu_{\mathrm{TLS}}}

\newcommand{\coupleresqb}{\ensuremath \nu_{\mathrm{res,QB}}}
\newcommand{\coupleresqbzero}{\ensuremath g_{\mathrm{res,QB}}}
\newcommand{\couplefluctqb}{\ensuremath \nu_{\mathrm{TLS,QB}}}
\newcommand{\couplefluctqbzero}{\ensuremath g_{\mathrm{TLS,QB}}}

\newcommand{\Irf}[1][]{\ensuremath I_{\mathrm{rf}#1}}
\newcommand{\Idc}[1][]{\ensuremath I_{\mathrm{dc}#1}}
\newcommand{\Icontrol}{\ensuremath I}
\newcommand{\nurf}{\ensuremath \nu_{\mathrm{rf}}}

\newcommand{\stddev}{\ensuremath {\mathrm{stddev}\,}}

\begin{document}
\title{A superconducting qubit as a quantum transformer routing
  entanglement between a microscopic quantum memory and a macroscopic
  resonator}

\author{Alexander Kemp}
\email{kemp@will.brl.ntt.co.jp}
\affiliation{NTT Basic Research Laboratories, NTT Corporation, 3-1 Morinosato Wakamiya Atsugi-shi, Kanagawa  243-0198 Japan}
\author{Shiro Saito}
\affiliation{NTT Basic Research Laboratories, NTT Corporation, 3-1 Morinosato Wakamiya Atsugi-shi, Kanagawa  243-0198 Japan}
\author{William J. Munro}
\affiliation{NTT Basic Research Laboratories, NTT Corporation, 3-1 Morinosato Wakamiya Atsugi-shi, Kanagawa  243-0198 Japan}
\affiliation{National Institute of Informatics 2-1-2 Hitotsubashi, Chiyoda-ku, Tokyo 101-8430, Japan}
\author{Kae Nemoto}
\affiliation{National Institute of Informatics 2-1-2 Hitotsubashi, Chiyoda-ku, Tokyo 101-8430, Japan}
\author{Kouichi Semba}
\email{semba@will.brl.ntt.co.jp}
\affiliation{NTT Basic Research Laboratories, NTT Corporation, 3-1 Morinosato Wakamiya Atsugi-shi, Kanagawa  243-0198 Japan}

\begin{abstract}
  We demonstrate experimentally the creation and measurement of an
  entangled state between a microscopic two level system and a
  macroscopic superconducting resonator where their indirect
  interaction is mediated by an artificial atom, a superconducting
  persistent current qubit (PCQB). We show that the microscopic two level
  system, formed by a defect in an oxide layer, exhibits an order of
  magnitude longer dephasing time than the PCQB, while the
  dephasing time of the entangled states between the microscopic two
  level system and macroscopic superconducting resonator is
  significantly longer than the dephasing time in the persistent
  current qubits. This demonstrates the possibility that a qubit of
  moderate coherence properties can be used in practice to address low
  decoherence quantum memories by connecting them to macroscopic
  circuit QED quantum buses, leading future important implications for
  quantum information processing tasks.
\end{abstract}
\date\today
\maketitle

The twentieth century saw the discovery of one of the most fundamental
and far reaching theories ever developed, quantum mechanics. Quantum
mechanics provides a set of principles describing physical reality at
the atomic level of nature and is critical to our understanding of how
atomic devices work. It has important implications for the processing
of information at this atomic level and in fact allows for a paradigm
shift to quantum information processing. Quantum mechanics provides a
fundamentally different computational model by employing features not
present in a classical world, most notably superposition and
entanglement. Such coherence properties however are not restricted to
the microscopic world. Quantum coherence in macroscopic objects have
been observed in a number of physical systems, which are usually
refered as macroscopic quantum coherence effects. These effects can be
most prominent in solid-state systems like engineered superconducting
electronic systems \cite{Clarke2008}. We can now design quantum
circuits on a controllable scale, making quantum mechanics available
as a technological resource. As an example, circuit quantum
electrodynamics experiments
\cite{Wallraff2004,Chiorescu2004,Johansson2006} have demonstrated the
coupling of artificial two-level systems (qubits) to single photons in
macroscopic superconducting resonators. Other researchers have
demonstrated  \cite{Majer2007,Sillanpaa2007} the use of such resonators as versatile quantum
buses \cite{Blais2004,Spiller2006,Kimble2008} to couple distant qubits, leading to experimental
demonstrations of quantum algorithms \cite{DiCarlo2009} and Bell violations \cite{Ansmann2009}. In
the longer term, with the promising developments in highly integrated
nanotechnologies, these are going to be important to construct
superconducting quantum information processors. However imperfections,
defects will degrade the performance of such devices. One particular
defect of recent interest are microscopic two level systems (TLS)
inside the barriers of Josephson junctions. These are individual
quantum objects \cite{Simmonds2004,Bertet2005,Lisenfeld2009,Kim2008,Lupascu2009,Simmonds2009}, usually acting as an inevitable noise source,
however they can also be used as proxies for engineered qubits \cite{Zagoskin2006},
their long coherence time making them a good proxy for quantum
memories, despite their limited controllability \cite{Neeley2008a}. This leads to the
natural question of how to access, manipulate and control TLSs. It is
essential for TLSs to be able to coherently interact with other
quantum objects within our system generating entanglement in between
the two systems. Of particular interest would be to create
entanglement between the TLS and the macroscopic resonator. However, neither of
these objects directly couple to one another and so would need to be
mediate by the superconducting qubits. In this letter, we introduce
and experimentally demonstrate a new design of entangling gate between
TLS and the resonator via superconducting qubits. Given the short
coherence time of the superconducting qubits, it is curious to know if
it would be only extremely short lived due to the coupling with the
short-lived superconducting qubits.

\begin{figure}
\includegraphics{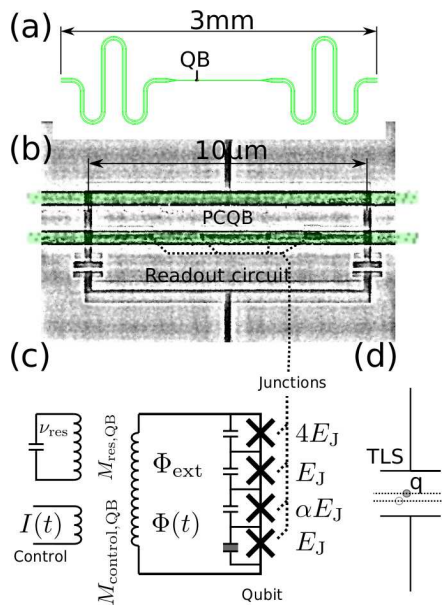}
\caption{\label{fig:schema} (a) Layout of the resonator (b) Sample
  micrograph; isolated resonator top layer marked up in color (c) Schematic of the
  qubit, the resonator and the control.  (d) A junction capacitor
  containing a TLS carrying a charge $q$ tunneling between two
  different positions. }
\end{figure}

This letter demonstrates the entanglement of a single microscopic TLS
and a engineered open ended superconducting coplanar stripline of
length $\lres=\unit{7.1}{mm}$ , the geometry of which is depicted in
Fig.~\ref{fig:schema}(a). These systems are off-resonant and coupled
weakly directly, but both interact strongly with a persistent current
qubit (PCQB) \cite{Mooij1999} in a four Josephson junction
configuration \cite{Bertet2004}, depicted in Fig.~\ref{fig:schema}(b),
when tuning the PCQB into resonance.  A mutual inductance between the
qubit and the resonator, created by placing the qubit directly on the
substrate and resonator segment in an metalization layer above,
isolated by a $Si O_2$layer couples the zero point fluctuation current
of the resonator to the qubit persistent current. The electric scheme
in Fig.~\ref{fig:schema}(c) contains the qubit loop including the
junctions with Josephson energies $\EJ$, $4 \EJ$ and $\alpha \EJ$,
where $\alpha=0.72$, forming the persistent current qubit, coupled to
the control line and the stripline resonators fundamental mode with a
wavelength $\lambda=2 \lres$, represented by a resonant circuit formed
by an inductance and a parallel capacitance. The qubit is coupled to
the TLS, by the electric field in one of the capacitances shunting a
junction, displayed in Fig.~\ref{fig:schema}(d). Such TLS are
imperfections in the oxide, first explicitly observed in phase qubits
\cite{Simmonds2004}. A trapped elementary charge tunnels between
stable positions, illustrated in \Fig{schema}(c). 

\begin{figure}
\includegraphics{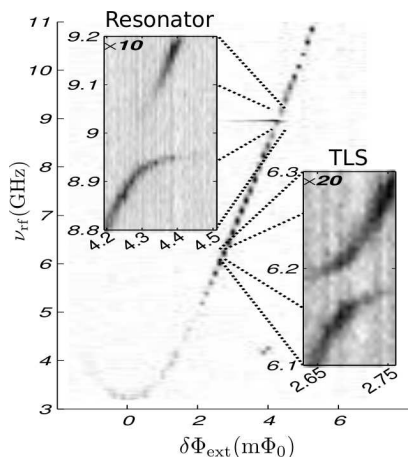}
\caption{\label{fig:spec} Spectrum of the qubit (density)
        showing the population of the qubits excited state as a
        function of field and freqeuncy.(Insets: higher resolution,
        enlarged).}
\end{figure}

The PCQB transition frequency $\nuqb$ is controlled during the
experiments by magnetic flux threading the PCQB loop, which is the sum
of a constant magnetic flux $\Phiext$, generated by a solenoid, and a
time-varying flux $\Phicontrol(t)$, induced by a current
$\Icontrol(t)$ through an on-chip control line coupled to the qubit by
a mutual inductance. The operating point is set to $\Phiext = 3
\Phizero/2 + \DeltaPhiext $ where $\Phizero = \planck / (\unit{2}{e})$
is the superconducting flux quantum and $\DeltaPhiext \ll \Phizero$.
At this bias two macroscopically distinct current states exist,
corresponding to a persistent current $\Ip\approx\pm\unit{300}{nA}$,
circulating clockwise or counterclockwise. The magnetic energy of
these two states corresponds to a frequency of $\epsilon(t) = \pm \Ip (\Phicontrol(t) +
\DeltaPhiext) / \planck$.  The Josephson junctions parameters generate a tunnel
element between the two states of $\Delta = \unit{3.2}{GHz}$.  The
PCQB Hamiltonian takes the form $\Hqb = \planck (\Delta \sigmaxQB
+ \epsilon(t) \sigmazQB) / 2$, where $\sigmaxQB$ and $\sigmazQB$ are the normal Pauli
matrices, so that the transition frequency is given by
$\nuqb=\sqrt{\Delta^2 + \epsilon^2(t)}$. The measured
spectrum after a long AC pulse $\Icontrol(t)=\Irf \cos(2 \pi \nurf t)$
with small amplitude $\Irf$ applied to the control line is plotted in
Fig.~\ref{fig:spec}, as a function of $\DeltaPhiext$, where the
frequency was swept for each magnetic field to acquire the spectrum of
the qubit. The qubit readout is achieved by a measurement DC-SQUID,
where after completing any qubit control sequence, in this case a long
rf pulse, a possible switching event of the measurement DC-SQUID is
recorded. Repeating this sequence $10000$ times enables to estimate
the qubit excitation probability.

While the dispersion relation on the coarse scale fits the qubit
dispersion relation, the insets show avoided crossings at the
microscopic TLS transition frequency $\nufluct = \unit{6.17}{GHz}$ and
the resonators fundamental freqeuncy $\nures = \unit{8.97}{GHz}$.  The
splitting of the avoided crossing at $\nures$ is $2 \coupleresqb =
\unit{112}{MHz}$, (corresponding to a coupling strength
$\coupleresqbzero = \unit{157}{MHz}$ in the canonical base of the
qubit). The TLS is presumably contained in a Josephson junction
barrier, indicated in Fig.~\ref{fig:schema}(b).  The coupling of a
charged TLS and a persistent current qubit was described in
\cite{Lupascu2009}, where the maximum possible coupling strength was
found to be $\couplefluctqbzero = \Delta \phim d/t $, where $\phim$ is
the phase difference associated with the persistent current states,
which is on the order of unity, $t \approx\unit{0.6}{nm}$ the oxide
layer thickness and $d \approx\unit{0.03}{nm}$ the distance between
the stable locations, so $\couplefluctqbzero \leq
\unit{200}{MHz}$. The experimentally observed splitting at $\nufluct $
is $2 \couplefluctqb = \unit{54}{MHz}$, corresponding to a coupling
strength of $\couplefluctqbzero = \unit{55}{MHz}$ in the qubits
canonical base. The TLS frequency $\nufluct$ fluctuated between the
acquisition of the spectrum in \Fig{spec} and the experiments
described below, which we attribute to a fluctuation of the
surrounding electric field \cite{Kim2008}.

\begin{figure}
\includegraphics{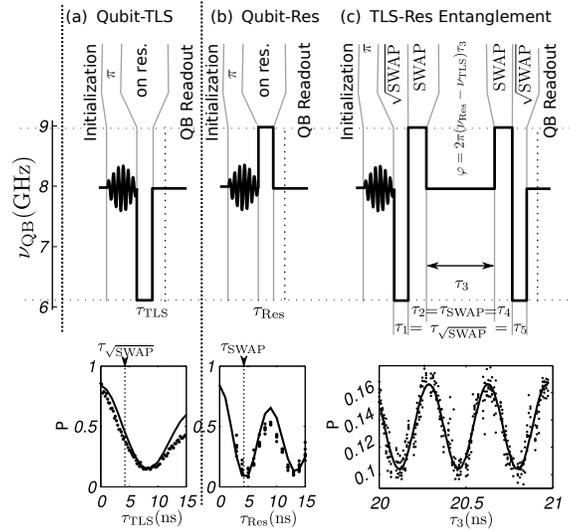}
\caption{\label{fig:pulse}
(a) Pulse sequence probing the coherent resonant
  interaction of the qubit and the TLS for a time
  $\tau_{\mathrm{TLS}}$. Bottom: measured qubit population (dots) and
  simulation result (line)\cite{Notesimu2}. The time required for a
  $\sqrt{\mathrm{SWAP}}$-operation is indicated.
(b) Pulse sequence probing the coherent
  resonant interaction of the qubit with the TLS by for a time
  $\tau_{\mathrm{res}}$. Bottom: measured qubit population (dots) and the
  simulation result (line). The time required for a
  $\mathrm{SWAP}$-operation is indicated.
(c) Pulse sequence for entanglement generation, phase evolution for a time $\tau_3$, and
  projecting the entangled state to the qubit state. Bottom: measured
  qubit population (dots) and fit (line) of a cosine, with the offset,
  phase, frequency and amplitude as free parameters.}
\end{figure}

During the time-domain experiments a constant external magnetic flux
$\DeltaPhiext \approx \unit{4}{m \Phizero}$ is applied, corresponding
to $\nuqb{} = \unit{7.92}{GHz} $. Before beginning any pulse sequence
the system is kept at $\Icontrol=0$ for approximately
$\unit{800}{\mu{}s}$. After this time the system has relaxed to its
thermodynamic equilibrium state, governed by the qubit energy $\planck
\nuqb$ and effective electronic qubit temperature of approximately $\unit{130}{mK}$,
corresponding to an exited state population of the qubit of about
$\unit{5}{\%}$, subsequently ignored in the discussion of the
sequences\cite{Notesimu}.
The system is then in the $\state{0,0,0}$ state, where the first
subspace denotes the PCQB state, the second denotes the TLS state,
and the third denotes the resonator state. Each of the sequences
depicted in Fig.~\ref{fig:pulse}(a)-(c) begins with a qubit spin flip
by a resonant $\pi$-pulse with a duration $\tau_0 \approx
\unit{4}{ns}$ and with an amplitude determined in a Rabi experiment,
preparing the system in the $\state{1,0,0}$ state.
The input current applied to the on-chip control line is defined by a
piecewise function\cite{Notepulsesequences} by the pulse lengths $\tau_i$ and the quasi-dc
pulse heights $\Idc[,i]$, corresponding to a qubit transition
frequency $\nuqb[,i]$. We represent the pulse sequences as shown in
Fig.~\ref{fig:pulse}(a)-(c) and model the system in the qubit and TLS
eigenbase by the approximate Hamiltonian $\Htotal = \planck (\nuqb(t)
(1/2) \sigmazQB + \nufluct (1/2) \sigmazTLS + \couplefluctqb \sigmaxQB
\otimes \sigmaxTLS + \nures (a^\dagger a + 1/2) + \,
\coupleresqb (\sigmamQB a+ \sigmapQB a^\dagger))$.  We first
characterize the qubit-TLS interaction as shown in \Fig{pulse}(a) by
bringing the excited qubit on resonance for a time
$\tau_{\mathrm{TLS}}$, allowing us to determine $\couplefluctqb$ and
the time $\tau_{\sqrt{\mathrm{SWAP}}}=1/(8 \couplefluctqb)$ needed for
performing a $\sqrt{\mathrm{SWAP}}$-operation.  After this operation
the TLS and the qubit are in the entangled state
$(\state{1,0,0}+\state{0,1,0})/\sqrt{2}$. This state is an entangled
state between the qubit and the TLS, named Bell-$\Phi^+$-state.  To
transfer the entanglement the resonator the qubit-resonator
interaction is characterize equivalently in \Fig{pulse}(b) and the time
$\tau_\mathrm{SWAP}=1/(4 \coupleresqb)$ corresponding to a
$\mathrm{SWAP}$-operation is determined, in which the qubit state and
the single-photon state are exchanged, making the state
$\state{0,0,1}$.

\begin{figure}
\includegraphics{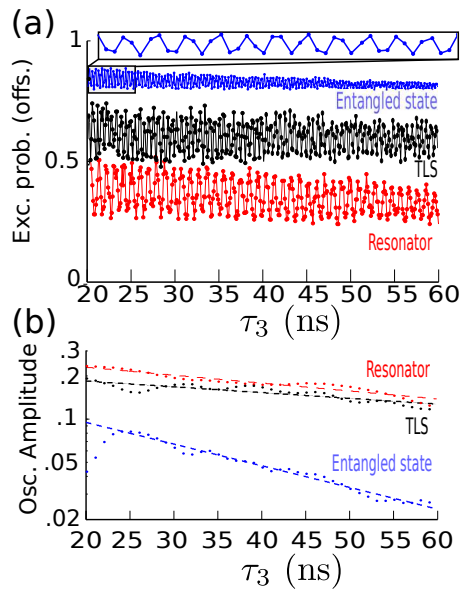}
\caption{\label{fig:demod} (a) Coherent oscillation of the qubit
  excitation acquired by Ramsey (memory) sequences on the TLS and the
  resonator, and the entanglement sequence in Fig.~2 as a function of
  time between the pulses.  (b) demodulated oscillation amplitudes;
  dashed lines indicate fits to exponential decays.}
\end{figure}

A combination of these two operations yields the pulse sequence in
\Fig{pulse}(c). The first quasi-dc pulse tunes the qubit in resonance
with the TLS for a time $\tau_1=\tau_{\sqrt{\mathrm{SWAP}}}$,
generating the entanglement and the second pulse brings qubit in
resonance with the resonator for a time $\tau_2=\tau_{\mathrm{SWAP}}$,
transferring the Bell state between the qubit and the TLS to a Bell
state between the TLS and the resonator $(\state{0,1,0}+\state{0,0,1})
/\sqrt2$. After this the qubit is taken off resonance for a time
$\tau_3$. During this period the
$(\state{0,1,0}+\state{0,0,1})/\sqrt2$ state evolves according to a
multi-spin Zeeman Hamiltonian as $(\exp(i 2 \pi \nufluct \tau_3 )
\state{ 0,1,0} + \exp(i 2 \pi \nures \tau_3) \state{0,0,1})/\sqrt2 =
\exp(i 2 \pi \nufluct \tau_3) (\state{(0,1,0)} + \exp(i
\varphi)\state{(0,0,1)})/\sqrt2$, where $\varphi=2 \pi \tau_3
(\nures-\nufluct)$.  As a result, the occupation oscillates between
the $\Phi^+, \varphi \equiv 0 (\mod 2\pi)$ state and the $\Phi^- ,
\varphi \equiv \pi (\mod 2\pi)$ state, so varying $\tau_3$ probes this
coherent phase evolution.  Bringing the qubit and the resonator in
resonance for a time $\tau_4= \tau_{\mathrm{SWAP}} $ swaps the
resonator state back to the qubit. The last pulse brings the qubit on
resonance with the TLS for $\tau_5= \tau_{\sqrt{\mathrm{SWAP}}}$
generating another $\sqrt{\mathrm{SWAP}}$-operation and projecting the
Bell-$\Phi$-states into the state
$\cos(\varphi)\state{1,0,0}+\sin(\varphi)\state{0,1,0}$, resulting in
an oscillating qubit excitation probability.  We plot the measured
excitation probability in \Fig{pulse}(c) as a function of $\tau_3$
between $\unit{20}{ns}$ and $\unit{21}{ns}$. Fitting using a robust
least-squares method yields frequency of
$\unit{2.935}{GHz}\pm\unit{22}{MHz}$, which matches the frequency
difference between the TLS and the resonator.  In \Fig{demod}(a) we show
the qubit excitation probability over $\tau_3$ ranging from
$\unit{20}{ns}$ to $\unit{60}{ns}$ in steps of $\unit{100}{ps}$. For
comparison, the results of Ramsey pulse sequences are shown, in which
memory sequence consisting of two SWAP operations with a varied time
delay is inserted, probing the coherence of the resonator and the
TLS. The Fourier transform (not shown) of the data in \Fig{demod}(a)
exhibits only a single dominant peak for each of the
signals. Demodulating the signal into amplitude and phase yields a
resonator frequency $\nures=\unit{8.97}{GHz}$, a TLS transition
frequency of $\nufluct=\unit{6.05}{GHz}$, and a entangled state
oscillation frequency of $\nu=\unit{2.92}{GHz}$, all to a precision
better than $\unit{10}{MHz}$ and fulfilling the relation
$\nures-\nufluct=\nu$.

Fitting the oscillation amplitudes, plotted in \Fig{demod}(b), to
exponential decays yields characteristic decay times and amplitudes
$A_0$ at $\tau_3=0$ for each of the coherent signals. Each of these
oscillation amplitudes is proportional to twice the absolute value
of the corresponding off-diagonal element in the density matrix. For
the entangled state, $A_0=0.2$, where the proportionality factor is
set by the transfer efficiency $\eta$ of the corresponding states in
the disentanglement gate.  Under the condition that only a single
photon is input into the system, which is fulfilled in the present
experiment\cite{Notesimu,Notesimu2}, the concurrence as a usual
entanglement measure \cite{Wootters1998} is given by twice the
absolute value of this density matrix element. Hence, regardless of
the numerical value of this factor, any finite oscillation amplitude
corresponds to a finite entanglement. In numerical simulations we
estimate $\eta \approx 0.6$ taking into account the Qubit dephasing
with a dephasing time of $\ttwoqb=\unit{14.8}{ns}$, and the
experimental pulse shaping limitations, yielding a concurrence for
$\tau_3=0$ of $C = A_0/\eta = 0.33$. This is a conservative
estimation, since the experimental $\eta$ is likely to be
lower. Independent of the absolute value of the concurrence, we can
probe its decay over time. A naive interpretation yields that the
decay can be described by a single decay rate, induced by relaxation
and dephasing of the off-diagonal entangled state density matrix
elements and should be the sum of the decay rate of the two subsystems
Ramsey decay rates. However, the memory dephasing times of
$\unit{73\pm8}{ns}$ for the resonator and $\unit{106\pm16}{ns}$ for
the TLS yield an expected characteristic decay time of the entangled
state of $\unit{43\pm6}{ns}$, which deviates from the experimental
value of $\unit{28\pm4}{ns}$. This suggests a mechanism for a
collective decoherence, which could be a direct coupling between the
resonator and the TLS or the indirect coupling via the qubit, which
participates in the resonator state due to the significant ratio of
$\coupleresqb/(\nuqb-\nures)$.

To conclude, we have shown an extremely heterogeneous quantum
experiment consisting of three effective qubits, in which a PCQB acts
as a transformer between a stripline resonator and a microscopic TLS
inside one of the qubits Josephson junctions.  We use this qubit to
generate, mediate and measure entanglement between the two other
systems.  The frequency of the entangled state matches the predicted
relations exactly, and the amplitude, equivalent to the concurrence,
decays over a timescale dominated by the relaxation of the
resonator. Overall we show that the qubit itself acts mainly as a
mediator and participates only weakly in the entangled states
dynamics, by observing a the entangled state coherence time
significantly longer than the qubit dephasing time.  This demonstrates
that a single active element, even with a short coherence time, can be
used as a mediator to generate entanglement between passive quantum
systems otherwise isolated from the environment.
\acknowledgments{{\em Acknowledgements}: We thank to K.~Kakuyanagi,
  F.~Deppe, H.~Yamaguchi, Y.~Tokura, X.~B.~Zhu for technical help and
  valuable comments.  This work was supported in part by Funding Program for World-Leading
Innovative R\&D on Science and Technology(FIRST), Scientific Research of
Specially Promoted Research \#18001002 by MEXT, and Grant-in-Aid for
Scientific Research (A) \#18201018 and \#22241025 by JSPS.}


\begin{thebibliography}{10}
\bibitem{Notesimu}
The excitation of the individual subsystems was estimated in another
  experiment; excitation of the TLS and the resonator were found to be
  negligible. Simulation results include these effects and the relaxation
  rates.
\bibitem{Notepulsesequences}
Pulse sequences were generated by digital-analog converters and subject to
  software based signal processing and heuristic optimization.
\end{thebibliography}
\end{document}